\documentclass[epj]{webofc}

\usepackage[varg]{txfonts}   
\woctitle{Mathematical Modeling and Computational Physics 2015}
\usepackage[english]{babel}
\usepackage{color} 
\def\adimH{\mathsf{K}} 
\def\alert{\emph} 
\newcommand{\AltG}[1]{\mathsf{A}_{#1}} 
\newcommand{\Aut}[1]{\mathrm{Aut}\vect{#1}} 
\newcommand{\barket}[1]{\left|#1\right\rangle} 
\newcommand{\brabar}[1]{\left\langle#1\right|} 
\def\bmat{\begin{pmatrix}}
\def\emat{\end{pmatrix}}
\def\C{\mathbb{C}} 
\newcommand{\cabs}[1]{\left|#1\right|} 
\def\cmath{\color{blue}}
\def\ctxt{\color{black}}
\newcommand{\CyclG}[1]{\Z_{#1}} 
\DeclareMathOperator{\diag}{diag}
\DeclareMathOperator{\e}{e}
\newcommand{\Frac}[2]{{#1/#2}}
\def\Entropy{\mathbf{S}} 
\def\Hspace{\mathcal{H}} 
\DeclareMathOperator{\im}{i}
\DeclareMathOperator{\idmat}{I}
\newcommand{\inner}[2]{\left\langle#1\!\mid\!#2\right\rangle} 
\newcommand{\IrrRep}[1]{\mathbf{#1}} 
\newcommand{\Math}[1]{$\cmath{}#1$} 
\newcommand{\MathEq}[1]{\begin{equation*}\cmath{#1}\end{equation*}}
\newcommand{\MathEqLab}[2]{\begin{equation}\cmath{#1}\label{#2}\end{equation}}

\newcommand{\Mtwo}[4]{\bmat#1&#2\\#3&#4\emat} 
\newcommand{\Mthree}[9]{\bmat#1&#2&#3\\
 #4&#5&#6\\
 #7&#8&#9\emat} 
\def\N{\mathbb{N}}
\def\natmod{\mathsf{H}} 
\def\NF{\mathcal{F}}
\DeclareMathOperator{\Obig}{O}
\DeclareMathOperator{\ord}{ord} 
\newcommand{\ordset}[1]{\left[#1\right]} 
\def\Prob{\mathbf{P}} 
\newcommand{\ProbBorn}[2]{\Prob\!\vect{#1,#2}} 
\def\period{\mathcal{K}} 
\DeclareMathOperator{\Period}{Period}
\newcommand{\projector}[1]{\Pi_{#1}}
\def\repq{\mathrm{U}} 
\newcommand{\set}[1]{\left\{#1\right\}} 

\DeclareMathOperator{\Sym}{Sym}
\def\Q{\mathbb{Q}} 
\def\regrep{\mathrm{P}} 
\newcommand{\SymG}[1]{\Sym\!\left(#1\right)} 
\def\R{\mathbb{R}} 
\DeclareMathOperator{\runi}{r}
\def\Time{\mathbf{T}} 
\DeclareMathOperator{\tr}{tr}
\def\transmatr{\mathrm{T}} 
\def\U{\mathsf{U}} 
\newcommand{\UG}[1]{\U\!\vect{#1}} 
\DeclareMathOperator{\Value}{val}
\newcommand{\vect}[1]{\left(#1\right)} 
\newcommand{\Vtwo}[2]{\bmat#1\\#2\emat} 
\newcommand{\Vthree}[3]{\bmat#1\\#2\\#3\emat} 
\def\wg{\mathsf{g}} 
\def\wG{\mathsf{G}} 
\def\wGN{\mathsf{M}} 
\def\ws{\omega} 
\def\wS{\Omega} 
\def\wSN{\mathsf{N}} 
\def\Z{\mathbb{Z}}
\def\Zenotime{\text{\Large\Math{\tau}}_{\!Z}}
\begin{document} 
\title{Combinatorial Approach to Modeling Quantum Systems}

\author{Vladimir~V.~Kornyak\inst{1}\fnsep\thanks{\email{kornyak@jinr.ru}}
}

\institute{Laboratory of Information Technologies,  Joint Institute for Nuclear Research, \\
141980 Dubna, Moscow Region, Russia
          }

\abstract{%
Using the fact that any linear representation of a group can be embedded into permutations, 
we propose a constructive description of quantum behavior that provides, in particular, a natural explanation of the appearance of complex numbers and unitarity in the formalism of quantum  mechanics.
In our approach, the quantum behavior can be explained by the fundamental impossibility to trace the identity of indistinguishable objects in their evolution.
Any observation only provides information about the invariant relations between such objects.
\par
The trajectory of a quantum system is a sequence of unitary evolutions interspersed with observations --- non-unitary projections.
We suggest a scheme to construct combinatorial models of quantum evolution. 
The principle of selection of the most likely trajectories in such models via the large numbers approximation leads in the continuum limit to the principle of least action with the appropriate Lagrangians and deterministic evolution equations.
}
\maketitle
\section{Introduction}
\label{intro}
Any continuous physical model is empirically equivalent to a certain finite model.
This is widely used in practice:  
solutions of differential equations by the finite difference method or by using truncated series are typical examples.
It is often believed that continuous models are ``more fundamental'' than discrete or finite models.
However, there are many indications that nature is fundamentally discrete at small (Planck) scales, and is possibly finite.%
\footnote{The total number of binary degrees of freedom in the Universe is about \Math{~ 10^{122}} as estimated via the holographic principle and the Bekenstein--Hawking formula.}
Moreover, description of physical systems by, e.g., differential equations can not be fundamental in principle, since it is based on approximations of the form
 \Math{f\vect{x}\approx{}f\vect{x_0}+\nabla{f\vect{x_0}}\Delta{}x}. 
In this paper we consider some approaches to constructing discrete combinatorial models of quantum evolution.
\par
The classical description of a \emph{reversible} dynamical system looks schematically as follows.
There are a set \Math{W} of states%
\footnote{The set \Math{W} often has the structure of a set of functions: \Math{W=\Sigma^X}, where \Math{X} is a \emph{space}, and \Math{\Sigma} is a set of \emph{local} states.}
and a group \Math{G_\mathrm{cl}\leq\SymG{W}} of transformations (bijections) of \Math{W}.
Evolutions of \Math{W} are described by sequences of group elements \Math{g_t\in{}G_\mathrm{cl}} parameterized by the {continuous} time \Math{t\in\Time=\ordset{t_a,t_b}\subseteq\R}.
The observables are functions \Math{h: W\rightarrow\R}.
\par
An arbitrary set \Math{W}  can be ``quantized'' by assigning numbers from a number system \Math{\NF} to the elements \Math{w\in{}W}, i.e., by interpreting \Math{W} as a basis of the module \Math{\NF^{\otimes{W}}}.
The quantum description of a dynamical system assumes that the module spanned by the set of classical states \Math{W} is a Hilbert space \Math{\Hspace_W} over the field of complex numbers, i.e., \Math{\NF=\C}.
The transformations \Math{g_t} and the observables \Math{h}  are replaced by {unitary}, \Math{U_t\in\Aut{\Hspace_W}},  and {Hermitian}, \Math{H}, operators on \Math{\Hspace_W}, respectively.
A constructive version of quantum description is reduced to the following:
\begin{itemize}
	\item 
time is \emph{discrete} and can be represented as a sequence of integers, typically \Math{\Time = \ordset{0,1,\ldots,T}};
	\item 
the set \Math{W} is finite and, respectively, the space \Math{\Hspace_W} is finite-dimensional; 
	\item 
the general unitary group \Math{\Aut{\Hspace_W}} is replaced by a finite group \Math{G};
	\item
the field \Math{\C}	is replaced by \Math{\period}th cyclotomic field \Math{\Q_\period}, where \Math{\period} depends on the structure of \Math{G};
	\item 
the evolution operators \Math{U_t} belong to a unitary representation of \Math{G} in the Hilbert space \Math{\Hspace_W} over \Math{\Q_\period}.
\end{itemize}
\par
It is clear that a single unitary evolution is \emph{not sufficient} for describing the physical reality.
Such evolution is nothing more than a physically trivial change of coordinates (a symmetry transformation). 
This means that observable values or relations, being invariant functions of states, do not change with time.
As an example, consider a unitary evolution of a pair of state vectors: \Math{\barket{\varphi_1}=U\barket{\varphi_0},} \Math{\barket{\psi_1}=U\barket{\psi_0}}.
For the scalar product we have \Math{\inner{\varphi_1}{\psi_1}=\brabar{\varphi_0}U^{-1}U\barket{\psi_0}\equiv\inner{\varphi_0}{\psi_0}}. 
There are two ways to obtain observable effects in the scenario of unitary evolution: 
(a) in quantum mechanics measurements are described by non-unitary operators --- projections into subspaces of the Hilbert space;
(b) in gauge theories collections of evolutions are considered, and comparing results of different evolutions can lead to observable effects (in the case of a non-trivial gauge holonomy).
\par
The role of observations in quantum mechanics is very important --- it is sometimes said that ``observation creates reality''.%
\footnote{The phrase is often attributed to John Archibald Wheeler.}
We pay special attention to the explicit inclusion of observations in the models of evolution.
While the states of a system are fixed in the moments of observation, there is no objective way to trace the identity of the states between observations.
In fact, all identifications --- i.e., parallel transports provided by the gauge group which describes symmetries of the states --- are possible.
This leads to a kind of fundamental indeterminism. 
To handle this indeterminism we need a way to describe statistically collections of parallel transports.
Then we can formulate the problem of finding trajectories with maximum probability that pass through a given sequence of states fixed by observations.
In a properly formulated model, the principle of selection of the most probable trajectories should reproduce in the continuum limit the principle of least action.

\section{Constructive description of quantum behavior}
\label{sec-quantum}
The transition from a continuous quantum problem to its constructive counterpart can be done by replacing a unitary group of evolution operators with some finite group.
To justify such a replacement \cite{KornyakPEPAN} one can use the fact from the theory of quantum computing that any unitary group contains a dense finitely generated subgroup.
This \emph{residually finite} \cite{Magnus} group has infinitely many finite homomorphic images. 
The infinite set of non-trivial homomorphisms allows to find a finite group that is empirically equivalent to the original unitary group in any particular problem.
\subsection{Permutations and natural quantum amplitudes}
\label{quantperm}
As it is well known, any representation of a finite group is a subrepresentation of some permutation representation.
Namely, a representation  \Math{\repq} of \Math{\wG} in a \Math{\adimH}-dimensional Hilbert space \Math{\Hspace_{\adimH}} can be embedded into a permutation representation \Math{\regrep} of \Math{\wG} in an \Math{\wSN}-dimensional Hilbert space \Math{\Hspace_{\wSN}}, where \Math{\wSN\geq\adimH}. 
The representation \Math{\regrep} is equivalent to an action of \Math{\wG} on a set of things \Math{\wS=\set{\ws_1,\ldots,\ws_\wSN}} by permutations.
If \Math{\adimH=\wSN} then \Math{\repq\cong\regrep}. Otherwise, if \Math{\adimH<\wSN}, the embedding has the structure
\MathEq{
\transmatr^{-1}\regrep\transmatr
=\Vtwo{
\left.
\begin{aligned}
\!\IrrRep{1}&\\[-2pt]
&\hspace{8pt}\mathrm{V}
\end{aligned}
\right\}\Hspace_{\wSN-\adimH}
}{
\left.
\hspace{29pt}
{\repq}
\right\}\Hspace_{\adimH}
},\hspace{10pt}
\Hspace_{\wSN} = \Hspace_{\wSN-\adimH}\oplus\Hspace_{\adimH}.
} 
Here \Math{\IrrRep{1}} is the trivial one-dimensional representation. It is a mandatory subrepresentation of any permutation representation. \Math{\mathrm{V}} is an optional subrepresentation.
We can treat the unitary evolutions of data in the spaces  \Math{\Hspace_{\adimH}} and \Math{\Hspace_{\wSN-\adimH}} {independently}, since both spaces are invariant subspaces of \Math{\Hspace_{\wSN}}. 
\par
The embedding into permutations provides a simple explanation of the presence of complex numbers and complex amplitudes in the formalism of quantum mechanics.
We interpret complex quantum amplitudes as 
projections onto invariant subspaces of vectors with natural components for a suitable permutation representation \cite{KornyakPEPAN,Kornyak12,Kornyak13a}.
It is natural to assign natural numbers --- multiplicities ---  to elements of the set \Math{\wS} on which the group \Math{\wG} acts by permutations.
The vector of multiplicities, 
\MathEq{\barket{n} = \Vthree{n_1}{\vdots}{n_{\wSN}},} 
is an element of the \emph{module} \Math{\natmod_\wSN = \N^\wSN}, where \Math{\N=\set{0,1,2,\ldots}} is the \emph{semiring} of natural numbers.
The permutation action defines the \emph{permutation representation} of \Math{\wG} in the module \Math{\natmod_\wSN}.
Using  the fact that all eigenvalues of any linear representation of a finite group are \emph{roots of unity}, we can turn the module \Math{\natmod_\wSN} into a Hilbert space \Math{\Hspace_{\wSN}}.
We denote by \Math{\N_\period} the \emph{semiring} formed by linear combinations of \Math{\period}th roots of unity with natural coefficients. 
The so-called \emph{conductor}  \Math{\period} is a divisor of the \emph{exponent}%
\footnote{The exponent of a group is defined as the least common multiple of the orders of its elements.}
 of \Math{\wG}.
In the case  \Math{\period>1} the semiring \Math{\N_\period} becomes a \emph{ring of cyclotomic integers}.
The introduction of the \emph{cyclotomic field} \Math{\Q_\period} as the field of fractions of the ring \Math{\N_\period} completes the conversion of the module \Math{\natmod_\wSN} into the Hilbert space \Math{\Hspace_{\wSN}}.
If \Math{\period>2}, then \Math{\Q_\period} is empirically equivalent to the field of complex numbers \Math{\C} in the sense that \Math{\Q_\period} is a dense subfield of \Math{\C}.  
\subsection{Measurements and the Born rule}\label{Born}
A quantum measurement is, in fact, a selection among all the possible state vectors that belong to a given subspace of a Hilbert space.
This subspace is specified by the experimental setup.
The probability to find a state vector in the subspace is described by the Born rule.
There have been many attempts to derive the Born rule from other physical assumptions --- the Schr\"{o}dinger equation, Bohmian mechanics, many-worlds interpretation, etc. 
However, the Gleason theorem \cite{Gleason} shows that the Born rule is a logical consequence of the very definition of a Hilbert space and has nothing to do with the laws of evolution of the physical systems.
\par
The Born rule expresses the probability to register a particle described by the amplitude \Math{\barket{\psi}} by an apparatus configured to select the amplitude \Math{\barket{\phi}} by the formula (in the case of pure states):
\MathEq{\ProbBorn{\phi}{\psi} = \frac{\cabs{\inner{\phi}{\psi}}^2}{\inner{\phi}{\phi}\inner{\psi}{\psi}}\equiv\frac{\brabar{\psi}\projector{\phi}\barket{\psi}}{\inner{\psi}{\psi}}\equiv\tr\vect{\projector{\phi}\projector{\psi}}\,,}
where \Math{\displaystyle\projector{a}=\frac{\barket{a}\!\brabar{a}}{\inner{a}{a}}} is the {projector} onto subspace spanned by \Math{\barket{a}}.\\
\textbf{Remark.}
In the ``finite'' background the only reasonable interpretation of probability is the \emph{frequency interpretation}: 
probability is the ratio of the number of ``favorable'' combinations to the total number of combinations. So we expect that \Math{\ProbBorn{\phi}{\psi}} must be a \emph{rational number} if everything is arranged correctly.
Thus, in our approach the usual \alert{non-constructive} contraposition
--- \alert{complex numbers} as intermediate values vs. \alert{real numbers} as observable values 
--- is replaced by the \alert{constructive} one --- \alert{irrationalities} vs. \alert{rationals}.
From the constructive point of view, there is no fundamental difference between irrationalities and constructive complex	numbers: both are elements of algebraic extensions.
\subsection{Illustration: constructive view of the Mach--Zehnder interferometer}
\label{MZI}
\begin{figure}[h]
\centering
\sidecaption
\includegraphics[width=0.4\textwidth]{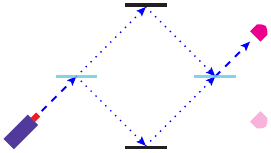}
\caption{Mach--Zehnder interferometer. Balanced setup: both beam splitters are \Math{50/50} and there is no phase shift between upper and lower paths.}
\label{MachZehnder} 
\end{figure}

The Mach--Zehnder interferometer is a simple but 
important example of a two-level quantum system. 
The device consists of a single-photon light source, beam splitters, mirrors and photon detectors (see Figure~\ref{MachZehnder}).
Consider a two-dimensional Hilbert space spanned by the two orthonormal basis vectors \Math{\barket{\nearrow}} --- ``right upward beams'', and \Math{\barket{\searrow}}  --- ``right downward beams''.
Then the \Math{50/50} beam splitter (i.e., a photon has equal probability of being reflected and transmitted) is described by the matrix 
\MathEqLab{S=\frac{1}{\sqrt{2}}\Mtwo{1}{\im}{\im}{1}\,.}{Smatrix}
The mirror matrix is \Math{~M=\Mtwo{0}{\im}{\im}{0}}\,. 
Notice that \Math{M=S^2}, and, on the other hand, \Math{S} can be expressed via \Math{M}  as an element of the group algebra: \Math{S=\frac{1}{\sqrt{2}}\vect{\idmat+M}}, where \Math{\idmat
} is the identity matrix. 
The scheme in the figure  implements the unitary evolution \Math{S\!MS\barket{\nearrow}=S^4\barket{\nearrow}=-\barket{\nearrow}}, 
which means that only the upper detector will register photons, the lower detector will always be inactive.
\par
This device is able to demonstrate many interesting features of the quantum behavior.
Consider, for example, the scheme of quantum \emph{interaction-free measurement} proposed by Elitzur and Vaidman \cite{ElitzurVaidman}.
The Penrose version of this example is called the \emph{bomb-testing problem}.
Suppose we have a collection of bombs, of which some are defective. 
The detonator of a good bomb causes explosion after absorbing a single photon.
The detonators of defective bombs reflect photons without any consequences.
Classically, the only way to verify that a bomb is good is to touch the detonator.
However, as shown in Figure~\ref{Bomb}, the quantum interference makes it possible to select  \Math{25\%} 
of good bombs without exploding them:
the signal of the lower detector ensures that the unexploded bomb is good.
\def\mpsz{0.46}
\def\grsz{0.6}
\begin{figure}[h]
\centering
\begin{minipage}{\mpsz\textwidth}
\centering 
\includegraphics[width=\grsz\textwidth]{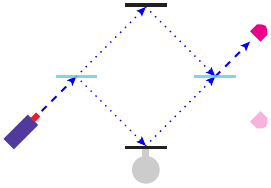}\\
{\small\Math{\barket{\nearrow}\xrightarrow{S\!MS}-\barket{\nearrow}~~\Prob=1}\\{}testing \alert{defective} bomb}
\end{minipage}
\begin{minipage}{\mpsz\textwidth}
\centering 
\includegraphics[width=\grsz\textwidth]{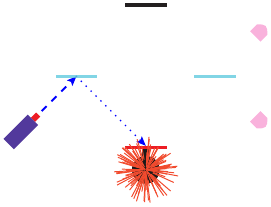}\\
{\small\Math{\barket{\nearrow}\xrightarrow{\projector{\searrow}S}\frac{\im}{\sqrt{2}}\barket{\searrow}~~\Prob=\frac{1}{2}}\\good bomb \alert{exploded}}
\end{minipage}
\\[5pt]
\begin{minipage}{\mpsz\textwidth}
\centering 
\includegraphics[width=\grsz\textwidth]{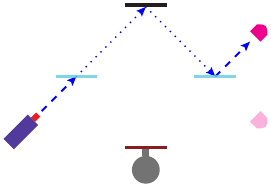}\\
{\small\Math{\barket{\nearrow}\xrightarrow{\projector{\nearrow}S\,M\,\projector{\nearrow}S}-\frac{1}{2}\barket{\nearrow}~~\Prob=\frac{1}{4}}\\bomb remains \alert{untested}}
\end{minipage}
\begin{minipage}{\mpsz\textwidth}
\centering 
\includegraphics[width=\grsz\textwidth]{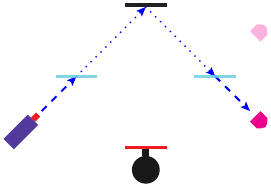}\\
{\small\Math{\barket{\nearrow}\xrightarrow{\projector{\searrow}S\,M\,\projector{\nearrow}S}\frac{\im}{2}\barket{\searrow}~~\Prob=\frac{1}{4}}\\bomb is \alert{good and intact}}
\end{minipage}
\caption{
Penrose bomb tester. \Math{\Prob} is the probability of a branch of evolution. \Math{\projector{a}} denotes the projector onto \Math{\barket{a}}.}
\label{Bomb} 
\end{figure}
\par
A slight modification of the scheme shown in Figure~\ref{MachZehnder}  allows us to implement any unitary operator \Math{U\in\UG{2}} by the Mach--Zehnder interferometer.
This is easily verified by direct calculation.
Since \Math{\dim\UG{2}=4}, we should add four parameters in a proper way.
For example, we can change the transparency of the beam splitter. Mathematically this means replacing the matrix \eqref{Smatrix} by another one of the form \Math{\alpha\idmat+\beta{\,}M}, where \Math{\cabs{\alpha}^2+\cabs{\beta}^2=1}.
Another possibility is to introduce \emph{phase shifters}. The phase shifter matrix related, e.g., to a ``right upward beam'' has the form \Math{\Mtwo{\e^{\im\omega}}{0}{0}{1}}.
Moreover, combining many Mach--Zehnder interferometers \cite{Zeilinger}, one can realize elements of any unitary group \Math{\UG{n}}.
\par
Since a ``mirror'' is the square of a ``beam splitter'', any unitary evolution in a sequence of balanced Mach--Zehnder interferometers can be described by degrees of \Math{S}.
The operator \Math{S} generates the cyclic group \Math{\CyclG{8}}.
The smallest degree faithful action of \Math{\CyclG{8}} is realized by permutations of \Math{8} objects.
Any of the four permutations, that generate \Math{\CyclG{8}} as a group of permutations, can be put in correspondence with the beam splitter, e.g., \Math{S\longleftrightarrow{g}=\vect{1,2,3,4,5,6,7,8}}. 
The generator \Math{g} can be represented by matrix \Math{P_{\!g}} acting in the module \Math{\N^8} that consists of the vectors with natural components: \MathEq{N=\vect{n_1,n_2,n_3,n_4,n_5,n_6,n_7,n_8}^T\in\N^8.}
To ``extract'' the beam splitter from the matrix \Math{P_{\!g}} we should extend the natural numbers by  \Math{8}th roots of unity --- the conductor \Math{\period=8} in this case.
Any \Math{8}th root of unity can be represented as a power of any of the four \emph{primitive} roots defined by the \emph{cyclotomic polynomial} \Math{\Phi_8\vect{\runi}=\runi^4+1}.
Let us denote by \Math{\N_8} the set of linear combinations of \Math{8}th roots of unity with {natural} coefficients. This is a ring since \Math{\period=8>1}.
The ring \Math{\N_8} is isomorphic to the ring of \Math{8}th cyclotomic integers.
In principle,  due to the projective nature of the quantum states, we could perform all calculations using only natural numbers and roots of unity.
But it is convenient to use also the \Math{8}th cyclotomic field, which we will denote by \Math{\Q_8}.
The field \Math{\Q_8} is the fraction field of the ring \Math{\N_8}.
\par
The matrix \Math{P_{\!g}} by a transformation \Math{T} over the field  \Math{\Q_8} can be reduced to the form
\MathEq{S_g=T^{-1}P_{\!g}T=\Mtwo{A}{0}{0}{S_{\!\runi}},}
where \Math{A=\diag\vect{1,-1,\runi^2,-\runi^2,\runi^3,-\runi}}, \Math{\runi} is a primitive \Math{8}th root of unity, and
\MathEqLab{\displaystyle{}S_{\!\runi}=\frac{1}{2}\Mtwo{\runi-\runi^3}{\runi+\runi^3}{\runi+\runi^3}{\runi-\runi^3}}{Srmatrix}
is the beam splitter matrix \Math{S} expressed in terms of the cyclotomic numbers.
Quantum amplitude of the Mach--Zehnder interferometer can be approximated by the projection of the natural vector \Math{N} into the ``splitter'' subspace:
\MathEqLab{\barket{\psi}=\Vtwo{\psi_1}{\psi_2}=\frac{1}{8}\Vtwo{-\runi{}^3\vect{n_1+n_3-n_5-n_7}+\vect{1-\runi{}^2}\vect{n_2-n_6}}{\runi{}\vect{n_1-n_3-n_5+n_7}+\vect{1+\runi{}^2}\vect{-n_4+n_8}}.}{N-to-psi}
It can be shown that expression \eqref{N-to-psi} can approximate with arbitrary precision any point on the Bloch sphere --- a standard representation of the complex projective line \Math{\C{}P^1}. 
\section{Combinatorial models of evolution}
\label{models}
Let us begin with some general 
considerations concerning the evolution of a probabilistic system subject to observations. 
The evolution of such a system can be described as follows.
We have a fundamental (``Planck'') time which is the sequence of integers: 
\MathEqLab{\Time = \ordset{0,1,\ldots,T}.}{Tfund}
There is also a sequence of ``times of observations''. 
For simplicity, we assume that the observation time is a subsequence of the fundamental time
\MathEqLab{\mathcal{T}=\ordset{t_0=0,\ldots,t_{i-1},t_{i},\ldots,t_{N}=T}}{Tobserv}
(otherwise we could assume that the times of observations are not determined exactly, e.g.,  they could be random variables with probability distributions localized within subintervals of the fundamental time).
Let \Math{W_{t_i}} denote the state of a system observed at the time \Math{t_i}, and 
\MathEqLab{W_{t_0}\rightarrow\cdots{\rightarrow}W_{t_{i-1}}{\rightarrow}W_{t_i}\rightarrow\cdots{\rightarrow}W_{t_N}}{traj}
denote a trajectory of the system.
Whereas the states \Math{W_{t_{i-1}}} and \Math{W_{t_i}} are fixed by observation, the transition between them can be described only probabilistically.
\par
The selection of the most probable trajectories is the main problem in the study of the evolution.
If we can specify \Math{\Prob_{W_{t_{i-1}}{\rightarrow}W_{t_{i}}}} --- the \emph{one-step transition probability} --- 
then the probability of trajectory \eqref{traj} can be calculated as the product 
\MathEqLab{\Prob_{W_{t_0}\rightarrow\cdots{\rightarrow}W_{t_{N}}}=\prod\limits_{i=1}^N\Prob_{W_{t_{i-1}}{\rightarrow}W_{t_{i}}}.}{Probtraj}
The inconvenience of dealing with the product of large number of multipliers can be eliminated by introducing the \emph{entropy}, which is defined as the logarithm of probability.
The transition to logarithms allows us to replace the products by sums.
On the other hand, taking the logarithm does not change the positions of the extrema of a function due to the monotonicity of the logarithm.
Thus, for searching the most likely trajectories we introduce the \emph{one-step entropy}
\MathEqLab{\Entropy_{W_{t_{i-1}}\rightarrow{}W_{t_{i}}}=\log\Prob_{W_{t_{i-1}}\rightarrow{}W_{t_{i}}}}{Entropy-one-step} 
and use instead of \eqref{Probtraj} the \emph{entropy of trajectory}:
\MathEqLab{\Entropy_{W_{t_{0}}\rightarrow\cdots\rightarrow{}W_{t_{N}}}=\sum\limits_{i=1}^N\Entropy_{W_{t_{i-1}}\rightarrow{}W_{t_{i}}}.}{Entropytraj}
\par 
The formulation of any dynamical model usually begins with postulating a Lagrangian.
However, it would be desirable to derive Lagrangians from more fundamental principles.
One can see that continuum approximations of \eqref{Entropy-one-step} and \eqref{Entropytraj} lead to the concepts of Lagrangian and action, respectively.
The reasoning is schematically the following.
The states \Math{W_{t_i}} are specified by sets of numerical parameters (coordinates) \Math{\mathbf{X}_{t_i}=\vect{X_{1,{t_i}}, X_{2,{t_i}},\ldots,X_{K,{t_i}}}}.
For a specific model one-step entropy \eqref{Entropy-one-step} can be calculated as a function of the coordinates: 
\Math{\Entropy_{W_{t_{i-1}}\rightarrow{}W_{t_{i}}}=S\!\vect{\mathbf{X}_{t_{i}},\Delta\mathbf{X}_{t_i}}}, where \Math{\Delta\mathbf{X}_{t_i}=\mathbf{X}_{t_{i}}-\mathbf{X}_{t_{i-1}}}.
Assuming that  \Math{\displaystyle{}N\rightarrow\infty,} \Math{\displaystyle{}t_i\!-\!t_{i-1}\rightarrow0} and embedding the sequence \Math{\mathbf{X}_{t_{i}}} into the continuous function \Math{\mathbf{X}\!\vect{t}},
we can represent the one-step entropy in the form \Math{S\!\vect{\mathbf{X}\!\vect{t_i}, \Delta\mathbf{X}\!\vect{t_i}}.} 
The second order Taylor approximation of this function has the form  \Math{\displaystyle{}S\approx{}A+b_{kk'}\vect{\Delta{X}_k\vect{t_i}-\Delta{X}_k^*\vect{t_i}}\vect{\Delta{X}_{k'}\vect{t_i}-\Delta{X}_{k'}^*\vect{t_i}}},
where \Math{\Delta\mathbf{X}^*\!\vect{t_i}} is the solution of the system of equations \Math{\displaystyle\frac{\partial{S}}{\partial\Delta\mathbf{X}\!\vect{t_i}}=0.}
Since the discrete time is a dimensionless counter, the differences can be approximated in the continuum limit by introducing derivatives, and we come to the Lagrangian
\MathEq{\displaystyle\mathcal{L}=A+B_{kk'}\vect{\frac{dX_k}{dt}-a_k}\vect{\frac{dX_{k'}}{dt}-a_{k'}},} where \Math{B_{kk'}} is a \emph{negative definite} quadratic form;~   
\Math{B_{kk'},~A} and \Math{a_{k}} depend on \Math{X_1\!\vect{t},} \Math{X_2\!\vect{t},\ldots,} \Math{X_K\!\vect{t}.}
The action \MathEq{\displaystyle\mathcal{S}=\int\!\mathcal{L}dt} is a continuum approximation of the entropy of trajectory \eqref{Entropytraj}, 
so the principle of least action can be treated as a continuous remnant of the principle of selection of the most likely trajectories.
\subsection{Example: extracting Lagrangian from combinatorics}
\label{Random-walk}
As an illustration of the above let us consider the one-dimensional random walk.
This model studies the statistics of sequences of positive (\Math{+1}) and negative (\Math{-1}) unit steps on the integer line \Math{\Z}.
Any statistical description is based on the concepts of \emph{microstates} and \emph{macrostates} --- the last can naturally be treated as equivalence classes of microstates \cite{Kornyak15}.
In this model, microstates are individual sequences of steps. The probability of a microstate consisting of \Math{k_+} positive and \Math{k_-} negative steps is equal to \Math{\alpha_+^{k_+}\alpha_-^{k_-}},
where \Math{\alpha_+} and \Math{\alpha_-} denote probabilities of single steps (\Math{\alpha_++\alpha_-=1}). 
The macrostates are defined by the equivalence relation: two sequences \Math{u} and \Math{v} are equivalent if \Math{k_+^u+k_-^u=k_+^v+k_-^v=t} and \Math{k_+^u-k_-^u=k_+^v-k_-^v=x}, 
i.e., both sequences have the same length \Math{t} and define the same point \Math{x} on \Math{\Z}.
The probability of an arbitrary microstate to belong to a given macrostate is described by the \emph{binomial distribution}, which in terms of the variables \Math{x} and \Math{t} takes the form
\MathEqLab{P\vect{x,t}=\frac{t!}{\vect{\frac{t+x}{2}}!\vect{\frac{t-x}{2}}!}\vect{\frac{1+v}{2}}^{\frac{t+x}{2}}\vect{\frac{1-v}{2}}^{\frac{t-x}{2}},}{binom} 
where \Math{v=\alpha_+-\alpha_-} is the ``drift velocity''.%
\footnote{It has been shown \cite{Knuth} that the velocity, defined in a similar way, i.e., as the difference of probabilities of steps in opposite directions, satisfies the relativistic velocity addition rule: \Math{w=\vect{u+v}/\vect{1+uv}}.}
Obviously, \Math{-1\leq{v}\leq1}.
\par
Let \Math{\ordset{x_0,\ldots,x_{i-1},x_{i},\ldots,x_N}} be a sequence of points (observed values) corresponding to the sequence of times of observations \eqref{Tobserv}.
We assume that the time differences \Math{\Delta{t}_i=t_i-t_{i-1}} are much larger than the unit of fundamental time \eqref{Tfund} but much less than the total time: \Math{1\ll\Delta{t}_i\ll{T}}.
Applying formula \eqref{binom} to \Math{i}th time interval we can write the one-step entropy:
\MathEq{\Entropy_{x_{i-1}\rightarrow{}x_i}=\ln\Delta{t}_{i}!-\ln\vect{\frac{\Delta{t}_{i}+\Delta{}x_i}{2}}!-\ln\vect{\frac{\Delta{t}_{i}-\Delta{}x_i}{2}}!
+\frac{\Delta{t}_{i}+\Delta{}x_i}{2}\ln\vect{\frac{1+v_i}{2}}+\frac{\Delta{t}_{i}-\Delta{}x_i}{2}\ln\vect{\frac{1-v_i}{2}},}
where \Math{\Delta{}x_i=x_i-x_{i-1}}, and \Math{v_i} denotes the drift velocity in the \Math{i}th interval.\\
Applying the Stirling approximation, \Math{\ln{}n!\approx{}n\ln{n}-n}, we have
\MathEqLab{\Entropy_{x_{i-1}\rightarrow{}x_i}\approx{}S_i=\Delta{t}_{i}\ln\Delta{t}_{i}-\frac{\Delta{t}_{i}+\Delta{}x_i}{2}\ln\vect{\frac{\Delta{t}_{i}+\Delta{}x_i}{1+v_i}}-\frac{\Delta{t}_{i}-\Delta{}x_i}{2}\ln\vect{\frac{\Delta{t}_{i}-\Delta{}x_i}{1-v_i}}.}{EStirling}
Solving the equation \Math{\partial{S_i}/\partial{\Delta{}x_i}=0} we obtain the stationary point: \Math{\Delta{}x_i^*=v_i\Delta{t}_{i}}. 
Replacing the sequences \Math{x_i,} \Math{v_i} by continuous functions \Math{x\vect{t},} \Math{v\vect{t}} and introducing the approximation \Math{\Delta{}x_i\approx\dot{x}\vect{t}\Delta{t}_{i}}
in the second order Taylor expansion of \eqref{EStirling} around the point \Math{\Delta{}x_i^*} we have finally
\MathEq{\displaystyle\Entropy_{x_{i-1}\rightarrow{}x_i}\approx-\frac{1}{2}\vect{\frac{\dot{x}\vect{t}-v}{\sqrt{1-v^2}}}^2\!\Delta{t}_{i}\,.}
Thus we come to the Lagrangian \Math{\displaystyle\mathcal{L}=\vect{\frac{\dot{x}\vect{t}-v}{\sqrt{1-v^2}}}^2}
with the corresponding Euler-Lagrange equation \MathEq{\displaystyle\frac{d}{dt}\frac{\partial\mathcal{L}}{\partial{\dot{x}}}-\frac{\partial\mathcal{L}}{\partial{x}}=0~\Longrightarrow~\ddot{x}\vect{1-v^2}+2\dot{x}v\frac{\partial{v}}{\partial{t}}-\vect{1+v^2}\frac{\partial{v}}{\partial{t}}=0\,.}

\subsection{Scheme for constructing models of quantum evolution}
\label{subsec-Qmodel}
The trajectory of a quantum system is a sequence of observations with unitary evolutions between them.
We propose a scheme to construct quantum models that combine unitary evolutions with observations.
The scheme assumes that transitions between observations are described by bunches of properly weighted unitary parallel transports.
The standard scheme of quantum mechanics with single unitary evolutions can be reproduced in our scheme by a special choice of weights.
But in our scheme such unique evolutions are assumed to be obtained as statistically dominant elements of the bunches.
\par
\begin{figure}[h]
\centering
\includegraphics[width=0.8\textwidth]{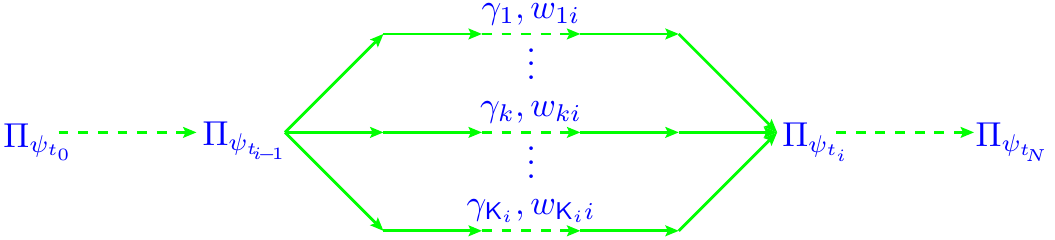}
\caption{Scheme of quantum evolution with observations}
\label{QModel} 
\end{figure}

We use the following notations
\begin{itemize}
	\item
\Math{\Hspace}: a Hilbert space;	
	\item
\Math{\projector{\psi_{t_0}},\ldots,\projector{\psi_{t_i}},\ldots,\projector{\psi_{t_N}}}: a sequence of observations,\\
where \Math{\projector{\psi_{t_i}}=\barket{\psi_{t_i}}\!\brabar{\psi_{t_i}}} is the projector that fixes \Math{\psi_{t_i}\in\Hspace} as the result of observation at the time \Math{t_i};
	\item
\Math{\Delta{t}_i=t_i-t_{i-1}}: the length of \Math{i}th time interval;
	\item
\Math{\wG=\set{\wg_1,\ldots,\wg_\wGN}}: a finite \emph{gauge group};	
	\item
\Math{\repq}: a unitary representation of \Math{\wG} in the space \Math{\Hspace};
	\item
\Math{\gamma=g_{1},\ldots,g_{\Delta{t}_i}}: a sequence of the length \Math{\Delta{t}_i}	of elements from \Math{\wG};
	\item
\Math{\Value\!\vect{\gamma}=\prod_{j=1}^{\Delta{t}_i}{g_{j}}\in\wG}: the (group) \emph{value} of the sequence \Math{\gamma} --- the parallel transport;
	\item
\Math{\Gamma_i=\set{\gamma_1,\ldots,\gamma_k,\ldots,\gamma_{\mathsf{K}_i}}}: an (arbitrary) \emph{enumeration} of the set of all sequences \Math{\gamma},\\
where \Math{\mathsf{K}_i\equiv\cabs{\Gamma_i}=\wGN^{\Delta{t}_i}} is  the total number of the sequences;
	\item
\Math{w_{ki}}: a \emph{non-negative weight} of \Math{k}th sequence (in \Math{i}th time interval).
\end{itemize}
With these notations we come to the scheme shown in  Figure~\ref{QModel}. 
The probability of transition from \Math{\psi_{t_{i-1}}} to \Math{\psi_{t_i}} is given by the formula
\MathEq{\Prob_{\psi_{t_{i-1}}\rightarrow\psi_{t_{i}}}=\sum\limits_{k=1}^{\mathsf{K}_i}\!\!w_{ki}\brabar{\varphi_{ki}}\projector{\psi_{t_i}}\barket{\varphi_{ki}}, 
\text{\ctxt~~where~~} \varphi_{ki}=\repq\!\vect{\Value\!\vect{\gamma_k}}\psi_{t_{i-1}}\,.}
The case of standard quantum mechanics with a single unitary evolution between observations is obtained in our scheme by selecting a sequence \Math{\gamma} formed by an element \Math{g\in\wG} repeated \Math{\Delta{t}_i} times.
The weight of the sequence \Math{\gamma} is set to \Math{1}, and the weights of all other sequences are equated to \Math{0}.
In other words,  the set of weights is the Kronecker delta on the set of sequences: \Math{w_{ki}=\delta_{\gamma,\gamma_k},~\gamma_k\in\Gamma_i}.
Introducing the Hamiltonian \Math{H=\im\ln\repq\!\vect{g}}, we can write the evolution in the usual form
\MathEq{U\equiv\repq\vect{g^{\Delta{t}_i}}=\e^{-\im{}H\vect{t_i-t_{i-1}}}.}
Since the notion of Hamiltonian stems from the principle of least action, it is natural to assume the existence of some mechanism of selecting sequences of the form \Math{g,g,\ldots,g} as dominant elements in the set of all sequences.
This requires a detailed analysis of the combinatorics of steps in fundamental time \eqref{Tfund} for particular models.

 \subsection{Dynamics of observed quantum system. Quantum Zeno effect and finite groups}
\label{subsec-Zeno}
Consider the issue concerning the connection between the quantum dynamics and the group properties of unitary evolution operators.
Namely, we consider the quantum Zeno effect for operators that belong to representations of finite groups.
\par
The ``quantum Zeno effect''
\footnote{This effect is also known under the name ``the Turing paradox''.}
 (see the review \cite{ZenoRev})
is a feature of the quantum dynamics, which is manifested in the fact that frequent measurements can stop (or slow down) the evolution of a system --- for example, inhibit decay of an unstable particle --- or force it to evolve in a prescribed way.
In the latter case, the phenomenon is called  the ``anti-Zeno effect''.
\par
Consider a quantum system that evolves from the initial (at \Math{t=0}) normalized pure state \Math{\barket{\psi_0}} under the action of the unitary operator \Math{U=\e^{-\im{H}t}}, where \Math{H} is the Hamiltonian.
The probability to find the system in the initial state at time \Math{t} is the following
\MathEqLab{{}p_H\vect{t}=\cabs{\left\langle{\psi_0}\cabs{\,\e^{-\im{H}t}}{\psi_0}\right\rangle}^2.}{ZenoProb}
The most important characteristics of any dynamical process are its temporal parameters.
For the quantum Zeno effect such a parameter is called the ``Zeno time'', denoted \Math{\tau_Z}.
It is determined from the short-time expansion of \eqref{ZenoProb}:
\MathEqLab{p_H\vect{t}=1-t^2/\tau_Z^2+\Obig\vect{t^4}.}{ZenoExp}
Calculation of \eqref{ZenoExp} shows that~~
\Math{\displaystyle\tau_Z^{-2}=\left\langle{\psi_0}\cabs{H^2}{\psi_0}\right\rangle-{\left\langle{\psi_0}\cabs{H^{\phantom{1}\!\!}}{\psi_0}\right\rangle}^2.} 
\par
Let us present the so-called \emph{Zeno dynamics} in the framework of scheme proposed in Section~\ref{subsec-Qmodel}.
We have here the sequence of observations
\Math{\projector{\psi_{t_0}},\projector{\psi_{t_1}},\ldots,\projector{\psi_{t_N}}},
each of which selects the same state \Math{\psi_0}, i.e., \Math{\psi_{t_0}=\psi_{t_1}=\cdots=\psi_{t_N}\equiv\psi_0}.
Assuming that \Math{t_0=0,~ t_N=T} and the times of observations are equidistant: \Math{t_i-t_{i-1}=T/N}, we can write, using \eqref{ZenoExp}, the approximation for the one-step transition probability 
\MathEq{\Prob_{\psi_{t_{i-1}}\rightarrow\psi_{t_{i}}}\approx1-\frac{1}{N^2}\vect{\frac{T}{\tau_Z}}^2}
with the corresponding approximation for the one-step entropy
\MathEq{\Entropy_{\psi_{t_{i-1}}\rightarrow{}\psi_{t_{i}}}\approx-\frac{1}{N^2}\vect{\frac{T}{\tau_Z}}^2.}
For the entropy of the trajectory we have
\MathEq{\Entropy_{\psi_{t_{0}}\rightarrow\cdots\rightarrow{}\psi_{t_{N}}}=\sum\limits_{i=1}^N\Entropy_{\psi_{t_{i-1}}\rightarrow{}\psi_{t_{i}}}\approx-\frac{1}{N}\vect{\frac{T}{\tau_Z}}^2\xrightarrow{N~\rightarrow~\infty}0}
and, respectively, for the probability of trajectory:\hspace{15pt}
\Math{\displaystyle\Prob_{\psi_{t_{0}}\rightarrow\cdots\rightarrow{}\psi_{t_{N}}}
\xrightarrow{N~\rightarrow~\infty}\e^0=1.}\\
This is precisely the essence of the Zeno effect.
\par  
Now assume that the evolution operator \Math{U} belongs to a representation of a finite group \Math{\wG}, i.e., \Math{U=\repq\vect{\wg},~\wg\in\wG}, and the time is the sequence of natural numbers: \Math{t={0,1,2,\ldots}}~.
A natural way to define the Zeno time in this case follows from the observation that the leading part of expansion \eqref{ZenoExp} vanishes at \Math{t=\tau_Z}.
By analogy we can define the \emph{natural Zeno time} \Math{\Zenotime}  as the first \Math{t\in\ordset{0,1,2,\ldots}} that provides minimum of the expression
\MathEqLab{p_U\vect{t}=\cabs{\left\langle{\psi_0}\cabs{U^{t}}{\psi_0}\right\rangle}^2.}{nZeno}
Obviously expression \eqref{nZeno} is either constant (namely, \Math{p_U\vect{t}=1}) or periodic. 
In the latter case its period is a divisor of the order of \Math{U}.
The \emph{order} of an element \Math{a} of a group is the smallest natural number \Math{n>0} such that \Math{a^n=e}, where \Math{e} denotes the identity element of the group.
The order of \Math{a} will be denoted \Math{\ord\vect{a}}. For the faithful representation we have \Math{\ord\vect{U}\equiv\ord\vect{\repq\vect{\wg}}=\ord\vect{\wg}}.
\par
Consider, for example, the ``Max-Zehnder'' representation \Math{\repq_{MZ}} of the group \Math{\CyclG{8}}, i.e., the ``beam splitter'' matrix \eqref{Smatrix} is taken as a generator of \Math{\CyclG{8}}.
Table~\ref{tabZ8} presents the Zeno times for all operators from the representation \Math{\repq_{MZ}}. 
We adopt the convention (motivated by formula \eqref{ZenoExp}) that \Math{\Zenotime=\infty} if probability \eqref{nZeno} is constant.
\begin{table}[h]
\centering
\caption{Zeno times for all operators from  \Math{\repq_{MZ}\vect{\CyclG{8}}}}
\label{tabZ8}
\begin{tabular}{c|c|c|c}
\Math{U=\repq_{MZ}\vect{\wg}} & \Math{\ord\!\vect{\wg}} &  \Math{\Period\!\vect{p_U\vect{t}}} &  \Math{\Zenotime}  \\\hline
\Math{S^0=\idmat} & \Math{1} &  \Math{p_U\vect{t}=1} &  \Math{\infty} \\\hline
\Math{S^4} & \Math{2} & \Math{p_U\vect{t}=1}  & \Math{\infty}  \\\hline
\Math{S^2=M, S^6} & \Math{4} &  \Math{2} &  \Math{1} \\\hline
\Math{S, S^3, S^5, S^7} & \Math{8} & \Math{4} & \Math{2}  
\end{tabular}
\end{table}
\par
\begin{figure}[h]
\centering
\includegraphics[width=\textwidth,height=0.31\textwidth]{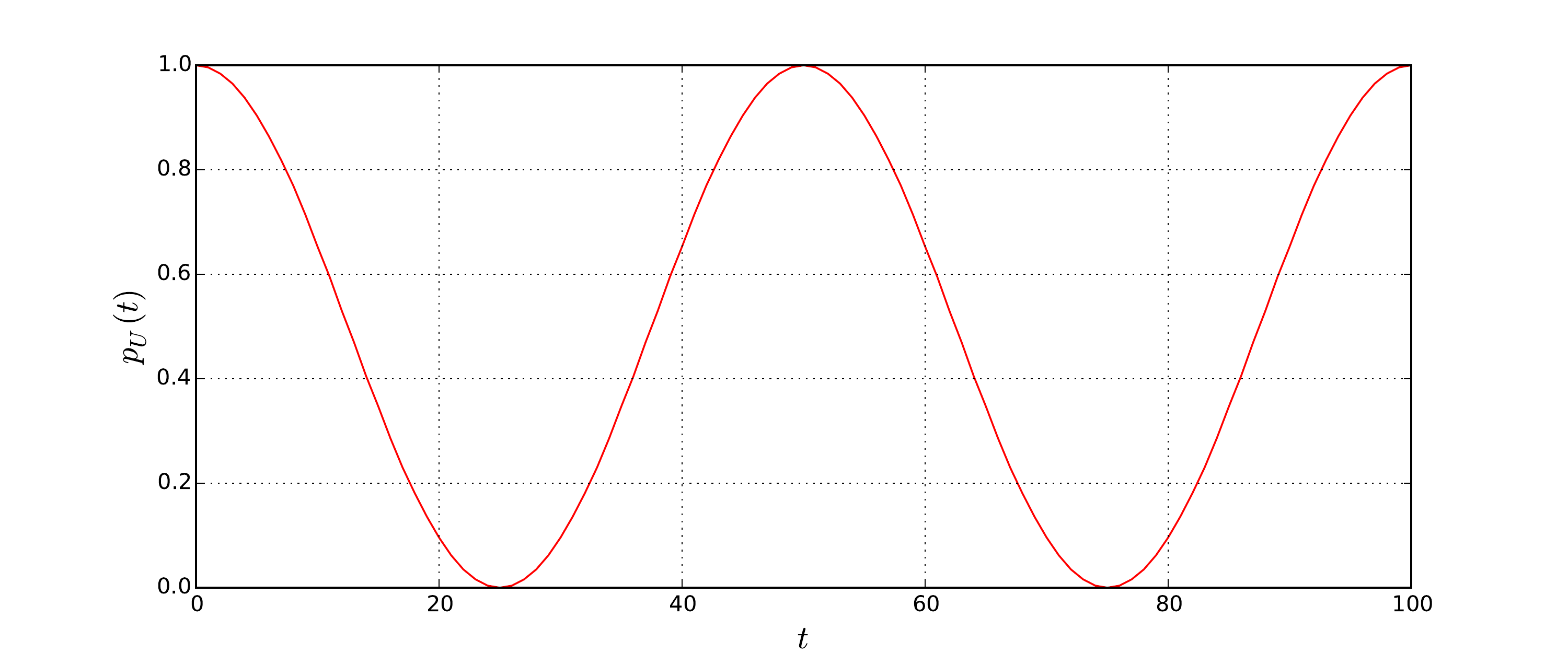}
\caption{Probability \Math{p_U\vect{t}} vs. time \Math{t} for the operator \Math{U=S_{\!\!100}\in\repq_{MZ}\vect{\CyclG{100}}}.}
\label{ZenoZ100} 
\end{figure}

The two-dimensional ``Max-Zehnder''  representation \Math{\repq_{MZ}} can be generalized to an arbitrary cyclic group \Math{\CyclG{N}} by replacing the  ``beam splitter'' matrix of the form \eqref{Srmatrix} with the unitary matrix 
\MathEq{S_{\!N}=\frac{1}{2}\Mtwo{\runi+\runi^{N-1}}{\runi-\runi^{N-1}}{\runi-\runi^{N-1}}{\runi+\runi^{N-1}}\,,}
where \Math{\runi} is an \Math{N}th primitive root of unity. 
Figure~\ref{ZenoZ100} shows the evolution of the probability to observe the initial state for the evolution operator \Math{S_{\!100}} in the time interval \Math{0\leq{t}\leq100}.
The quadratic short-time behavior, described by the formula \eqref{ZenoExp}, is clearly visible in the figure. The Zeno time in this example is \Math{\Zenotime=25}.
\par
\begin{figure}[h]
\centering
\includegraphics[width=\textwidth,height=0.31\textwidth]{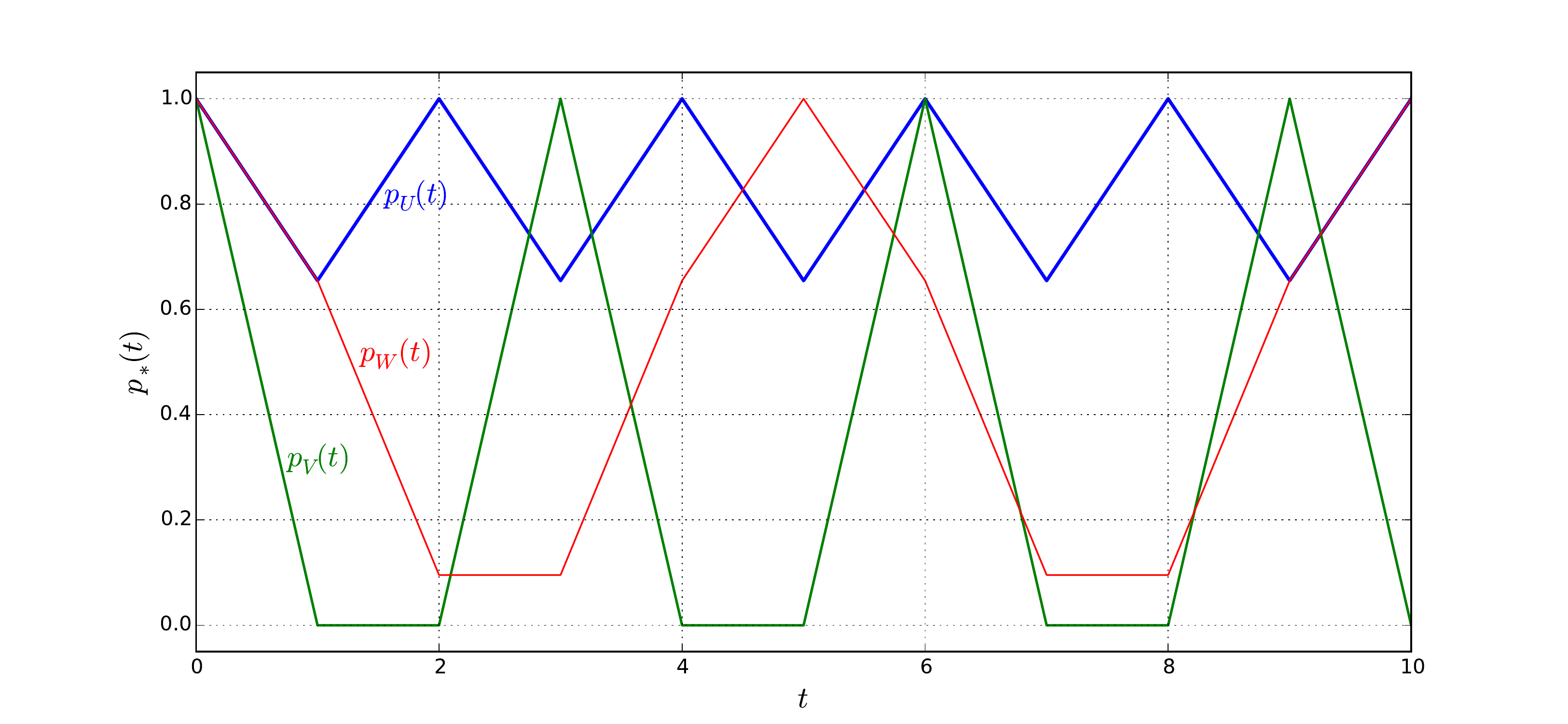}
\caption{Zeno dynamics in the representation \Math{\IrrRep{3'}} of the group \Math{\AltG{5}} for unitary operators \eqref{operators}.}
\label{ZenoA5} 
\end{figure}

As a non-commutative example, consider the icosahedral group \Math{\AltG{5}} --- the smallest (\Math{\cabs{\AltG{5}}=60}) non-commutative simple group.
It has applications for model building in the particle physics, especially in issues beyond the standard model, such as the flavor physics \cite{EverettStuart}.
The non-trivial elements of \Math{\AltG{5}} have orders \Math{2}, \Math{3} and \Math{5}.
The irreducible representations of \Math{\AltG{5}} are: one trivial singlet, \Math{\IrrRep{1}}, two triplets, \Math{\IrrRep{3}} and \Math{\IrrRep{3'}}, one quartet, \Math{\IrrRep{4}}, and one quintet, \Math{\IrrRep{5}}.
Figure \ref{ZenoA5} shows the evolution of ``Zeno probabilities'' for the following matrices  of orders \Math{2}, \Math{3} and \Math{5}, respectively,  
\MathEqLab{U=\frac{1}{2}\Mthree{-\phi}{\Frac{1}{\phi}}{1}{\Frac{1}{\phi}}{-1}{\phi}{1}{\phi}{\Frac{1}{\phi}},~V=\Mthree{0}{0}{1}{1}{0}{0}{0}{1}{0},~W=\frac{1}{2}\Mthree{-\phi}{-\Frac{1}{\phi}}{1}{\Frac{1}{\phi}}{1}{\phi}{-1}{\phi}{-\Frac{1}{\phi}}\,,}{operators}
where \Math{\phi=\frac{1+\sqrt{5}}{2}} is the ``golden ratio''.
To write these matrices, we added an 
element of order \Math{3} (the simplest among randomly selected) to the generators of orders \Math{2} and \Math{5} proposed in \cite{Shirai} for the representation \Math{\IrrRep{3'}}.
\section{Summary}
\label{Summary} 
\begin{enumerate}
\item 
We adhere to the idea of empirical universality of discrete, more specifically, finite models for describing physical reality.
In other words, any continuous model can be replaced by a finite model that fit the same observable behavior.
\item
This idea, in application to quantum problems, means that unitary groups of evolution operators can be replaced by unitary representations of finite groups.
\item
The mathematical fact that any representation of a finite group can be embedded in a permutation representation allows to approximate, with arbitrary precision, quantum amplitudes by projections of vectors with natural components.
The complex components of these projections are combinations of natural numbers and roots of unity.
\item
To illustrate the content of the article, we have used the Mach-Zehnder interferometer --- a simple but important example of a two-level quantum system with rich behavior.
\item
We propose a scheme for constructing quantum models.
Taking into account that a single unitary evolution, being a simple change of coordinates,  is not sufficient to describe  physical phenomena,
the scheme involves sequences of observations with  bunches of unitary parallel transports between the observations.
\item
The principle of selection of the most probable trajectories in such models via the large numbers approximation leads in the continuum limit to the principle of least action with appropriate Lagrangians and deterministic evolution equations.
\item
To look at  the connection between quantum dynamics and the group properties of unitary evolution operators, we have considered the quantum Zeno effect in the context of our approach.
\end{enumerate}
\begin{acknowledgement}
The work is supported in part by the Ministry of Education and Science of the Russian Federation (grant 3003.2014.2) and the Russian Foundation for Basic Research (grant 13-01-00668).
\end{acknowledgement}

\end{document}